# Reservoir Computing using Stochastic p-Bits

Samiran Ganguly[1], Kerem Y. Camsari[2], and Avik W. Ghosh[1]
[1]Dept. of Electrical and Computer Engineering, University of Virginia, Charlottesville, VA 22904, USA
[2]School of Electrical and Computer Engineering, Purdue University, West Lafayette, IN 47907, USA
Emails: sganguly@virginia.edu, kcamsari@purdue.edu, ag7rq@virginia.edu

*Abstract*—We present a general hardware framework for building networks that directly implement Reservoir Computing, a popular software method for implementing and training Recurrent Neural Networks and are particularly suited for temporal inferencing and pattern recognition. We provide a specific example of a candidate hardware unit based on a combination of soft-magnets, spin-orbit materials and CMOS transistors that can implement these networks. Efficient non von-Neumann hardware implementation of reservoir computers can open up a pathway for integration of temporal Neural Networks in a wide variety of emerging systems such as Internet of Things (IoTs), industrial controls, bio- and photo-sensors, and self-driving automotives.

## I. Introduction

Temporal inferencing and learning is believed to be the next frontier in the discipline of Artificial Intelligence. Along with development of algorithms for temporal machine learning and development of hardware that can implement these algorithms efficiently and locally on chip have the potential to revolutionize the rapidly emerging era of smart-sensors, self-driving automotives, and IoTs. Hardware designed specifically to implement these algorithms can open a pathway for eventual real-time autonomous *in-situ* cognition at low energy costs as well as independence from a cloud based computing backed that the present day systems need, with major impact on security, net bandwidth and network disruption resilience.

Reservoir Computing (RC) [1,2] is a prominent software method that has shown promise in tasks such as learning and predicting a time series, channel equalization[3], acoustic data processing[4], video surveillance[5] etc. RC is advantageous due to the ease of implementation as it avoids the difficult problem of backpropagation through time[6] in Recurrent Neural Networks and is, therefore, an attractive approach to the problem, particularly for a hardware implementation.

In this work, using simulation-based studies, we show that RC methods, in particular Echo-State Networks (ESNs) can be efficiently implemented using devices such as soft-magnets that work in the stochastic regime with tunable and predictable noise profile. We focus on the recently proposed probabilistic (p)-bits[7] due to the one-to-one mapping between the dynamics of the networks of p-bits and ESNs. We show p-bit networks can successfully learn, and predict the output of a deterministic but chaotic time series, work as temporal auto-encoders, and can extract a signal from a noisy channel. These simulations show that such stochastic networks can have wide-ranging applications in signal processing, smart-sensors, temporal information compression, and wireless communication.

## II. Reservoir Computing

Reservoir Computing, including its two versions. the ESN and Liquid State Machines (LSMs) are inspired by the principles of biological neural networks[8,9]. In these networks (fig.1), the computation is performed by a collection of loosely, sparsely, and randomly coupled non-linear units with feedbacks (recurrence) which: (a) provides a huge expansion of the phase space, making the task of signal classification easier, and (b) gives rise to memory states in the network[10]. The nodes of the network are "leaky"; therefore, this memory is a short term and fading one, thus the name "echo-state" networks. The system of equations is given by:

$$\frac{dx}{dt} = -\eta x + k\,tanh(W^{in}u + W^{fb}y + W^{self}x) \quad (1)$$

$$y = W^{out}x \quad (2)$$

where x is the network state vector, u is the input vector, y is the output vector, η and k are constants, W's are interconnection matrices. In these networks, excitations/activations of the nodes are temporally correlated with the past and the present sequences from the input stream, while the stochasticity enables a short-term fading memory, which then enables efficient signal pipelining. We can use a linear combination of these activations to learn a time series data, classification, and in non-linear filtering tasks. The only connections adjusted during the training are the reservoir-to-output ones ($W^{out}$) using techniques such as ridge regression.

Reservoir Computers have previously been demonstrated on a wide variety of systems[11,12,13,14]. In this paper, we demonstrate it in compact hardware using p-bits utilizing their one-to-one correspondence with ESN nodes. These p-bits have controllable stochastic transfer functions, with built in switching delay (fig. 2). The controllable stochasticity provides the leakiness necessary for RC networks, and allows the system to get out of local minima in the energy landscape and anneal to the ground state. While stochastic transfer functions are found in a wide variety of emerging materials, such as phase-change, metal-insulator transition, multiferroics, or possibly even CMOS working in deep sub-threshold, in this work we focus on a cell design incorporating super-paramagnets (SPM) and high spin-orbit materials that offer natural current addition through the Giant Spin Hall Effect (GSHE).

## III. Soft-Magnet p-Bits

We introduce the notion of a p-bit and describe one possible hardware implementation of probabilistic, or p-bits[7]. The stochastic behavior of a p-bit can be compactly expressed as:

$$m = sgn[rand(-1,+1) + tanh(I_i)] \quad 3$$

where at each normalized time step τ (=0,1,2...), the bit makes a random transition to a new state +1 or -1 in the presence of a normalized biasing current $I_i$. The time step is assumed as long enough for past memory of the bit to be lost. In the absence of any bias current, the magnetization that is produced is a +1 and -1 with equal probability and a positive or negative bias current tunes this probability between 0 and 1 (Fig. 3).

The long-time average of the magnetization m (<m>) has a sigmoidal [~ $tanh(I_i)$] characteristics and this intrinsically maps to the physics of low-barrier nano-magnets. Therefore, a possible CMOS-assisted implementation of a p-bit that can be interconnected in a hardware implementation is a *low-barrier, circular nanomagnet* [7,15] placed on top of a heavy metal with GSHE (Fig 3.a.) whose charge current can induce a spin-current to pin the magnetization of the circular magnet in the plane of the magnet. The GSHE, being a current based component, allows for *addition of input currents naturally without any extra hardware*. A Magnetic Tunnel Junction (MTJ) using a small read current senses the magnetization. The voltage output is amplified and isolated by a set of CMOS inverters that are biased in the middle, relating input currents to output voltage spikes (between 0 and $V_{DD}$) that can be tuned by the charge current flowing in the GSHE, in the presence of thermal noise on the magnets (Fig 3.b).

The key feature such p-bits exhibit is that their *stochasticity is tunable by an input current*, hence they can be interconnected to demonstrate large scale, robust correlations (Fig 3.c.).

## IV. RESERVOIR DYNAMICS AND EXAMPLE SYSTEMS

We setup p-bits networks of various sizes (N=10, 20, 200, 500, 100) with $W^{in}$ and $W^{self}$ connection matrices whose individual entries are chose randomly. One of the hyperparameters of RC is the spectral radius (maximal absolute eigen value) of the $W^{self}$, and we scale it to be less than one, as this has been found to create desirable dynamical (echo-state) properties in the network. Stronger connections can make the reservoir stiff and give rise to spurious dynamical states[16].

For a N = 500 node network (fig. 4.a) excited with a time varying analog signal, we observe the activations of a few nodes of the network. We notice that these activations are correlated with the input signal variation, but different from each other (fig. 4.b.), as if each node remembers different parts of the input data stream. These excitations are combined (fig. 4.c) using a linear sum unit and the $W^{out}$ adjusted to perform various learning tasks.

We setup following tasks popular in RC literature for our network as proof-of-concept demonstrations:

### A. Chaotic Mackey-Glass System Predictor

Mackey-Glass (MG) equation[17] is a time series generator with periodic but subtly chaotic characteristics having origins in biology/psychology. The equation is given by:

$$\frac{dx}{dt} = \beta \frac{x(t-\tau)}{1+x(t-\tau)^n} - \gamma x, \quad \gamma, \beta, n > 0 \quad 4$$

We train out networks (fig. 5.a) on a MGS system with training datastream/signal of 5000 time steps using ridge regression to work as unity signal follower (the simplest test case: Identity function), and then tested on a different sequence of signal from MGS (fig. 5.b). We found that the network could reproduce the test signal faithfully with normalized mean square errors ranging from $1 \times 10^{-3}$ to $5 \times 10^{-4}$ for N = 10 to 1000 (fig. 5.c). We then use the network as a *temporal auto-encoder*, in which the network attempts to reproduce the test signal looking only at its previously self-generated output. We found that for small N the networks fails but starts to generate better results with larger N (fig. 5. d). This shows the possibility of creating temporal generative models using these networks.

### B. Channel Equalizer

We then turn our attention to channel equalization[18,19], a task of practical importance in wireless communication. The network (fig. 6.a) extracts the original signal d(t) from a noisy channel q(t) with non-linear and time shift distortions to generate an input sequence u(t) from d(t) and given by:

$$u(t) = \sum_{n=1}^{3} A_n \sum_{\tau=-2}^{7} B_k [d(t-\tau)]^n + rnd(-c,c) \quad 5$$

We train the network to learn to "invert" q(t) by using d(t) as the learning target and u(t) as the training signal. After training, we test the network and we find that for moderate sized networks (N=100), the symbol error rate (normalized difference between d(t) and output y(t) over a range of t), *SER can theoretically be zero* (fig. 6.b). In a practical circuit, SER will be limited by the values of elements in $W^{out}$ matrix that can be implemented due to component fabrication constraints.

## V. DISSIPATION IN P-BIT BASED RESERVOIRS

For any new nano-device/circuit proposal, it is critical to provide estimate of energy dissipation for its evaluation as a possible technology of interest. While metrics with wide consensus (e.g. E-D product) are available for Boolean gates, it is not always clear what should be the equivalent figure of merit for the kind of computation we are proposing, considering that implementation details, circuit design, and bias point choices can have a strong impact on such metrics. Therefore, we merely provide estimates for the $I^2R$ power dissipated in various components of a single p-bit under moderate overdrive conditions[20] as listed in the tab. 1. a, b.

We build the buffer using two inverters in series and can show[21] using a predictive model for 14nm FinFETs[22] that the dynamic dissipation in the buffer is about 150 μW. Further this sets the bias voltage that can be used in the circuit (here Vdd = 0.8V). As a result, the dissipation in the reader unit is about 20 μW, while in our case GSHE writer dissipates a negligible power of about 0.2 μW for a net write current of 15 μA. These calculations show that for a *dissipative cost of just over two CMOS inverters we can create a compact RC node with very high functionality-to-power ratio*. Using low power transistors can reduce these numbers even more, making it an attractive candidate for building temporal Neural Networks.

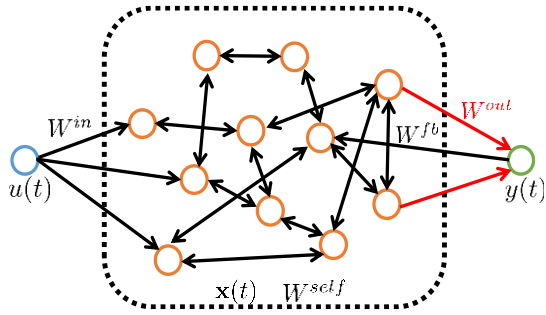

**Fig. 1. General schematic of a reservoir computer.** The reservoir is a network of sparse and randomly connected neurons (black arrows) captured by the weight matrix $W^{self}$. The reservoir $\mathbf{x}(t)$ (connected by $W^{in}$) processes the input bit stream $u(t)$ and produces the output bit stream $y(t)$ (feedback to reservoir by $W^{fb}$) as a response. The reservoir-to-output connection $W^{out}$ (red arrows) is the only layer trained in RC.

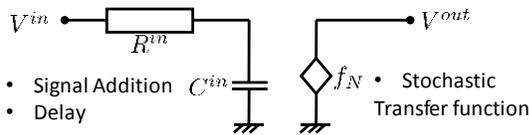

- Signal Addition
- Delay
- Stochastic Transfer function

**Fig. 2. Generic circuit representation of a reservoir node element.** Input side provides a delay element ($\tau = R^{in}C^{in}$), and an adder for time varying input signals, while the output side is a controlled voltage/current generator that produces a time varying output signal given by a transfer function which is a random number generator with controllable statistical properties (mean and variance).

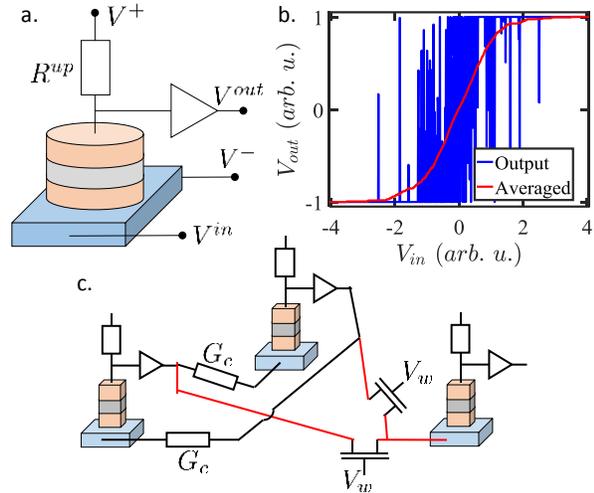

**Fig. 3. Hardware implementation of Reservoir Computing. a.** A possible implementation of RC nodes built using super-paramagnets driven by spin-orbit torques. The readout is through an MTJ, which forms a voltage divider pair with the pull up resistor with symmetric mid-point bias (see ref. 7 for details). **b.** The output of this unit is a random number generator whose statistical properties are controlled by the net signal at the input node. **c.** The nodes can then be connected with either conductors ($G_c$) of randomly chosen magnitudes (for the reservoir), or for the $W^{out}$ connections (shown in red) using a programmable mechanism, a few examples of which are floating gates on transistors operating in subthreshold regime ($V_w < V_t$), multiferroics, or phase-change materials/memristors.

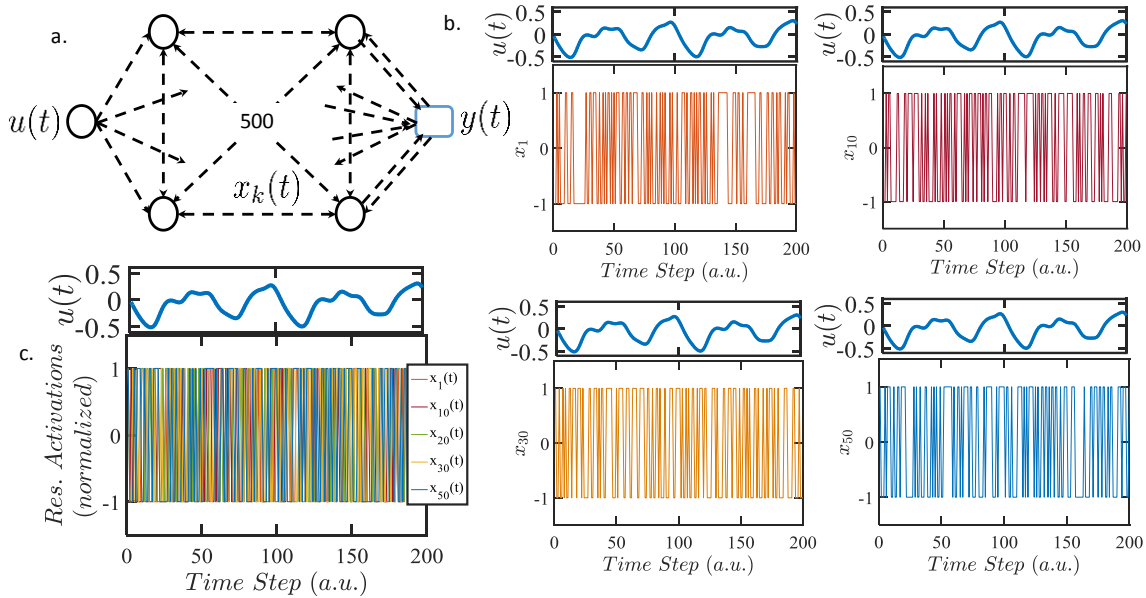

**Fig. 4. Example of dynamical activations in reservoir. a.** A 500 node RC with randomly connected weight matrices. **b.** Example of dynamical activations on nodes numbered 1, 10, 30, and 50. Each node activation is subtly different but correlated with input signal $u(t)$ and from $W^{self}$ connections with other nodes. **c.** All the activations put together show temporal correlations with the input signal $u(t)$ which is then used to construct $y(t)$ through adjustable matrix $W^{out}$ calculated using ridge regression on training signal.

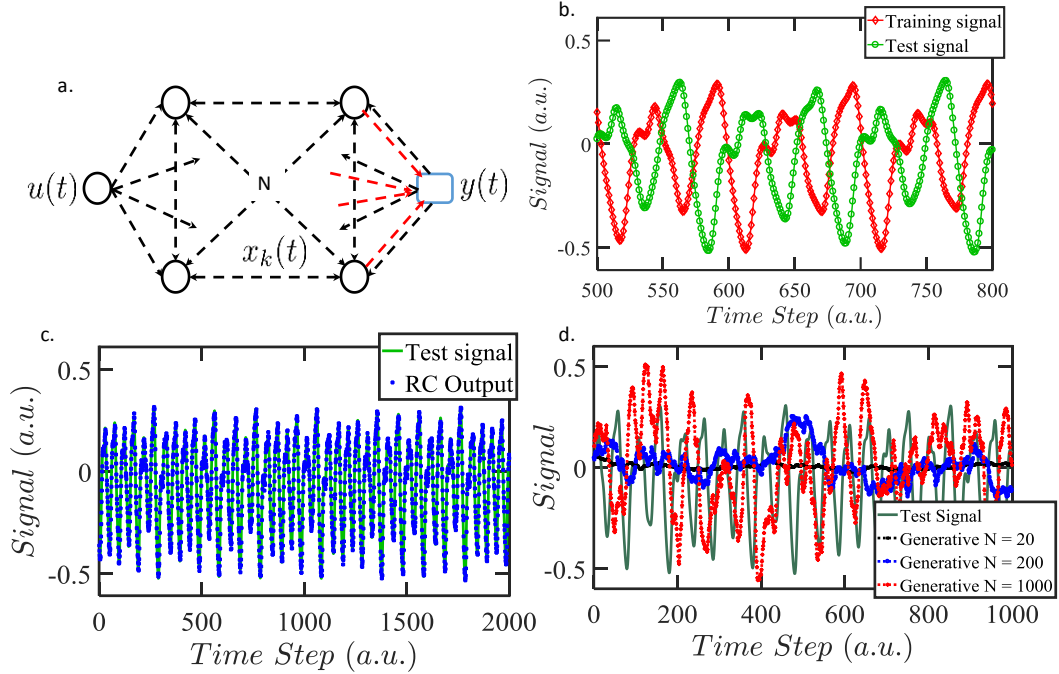

**Fig. 5. Mackay-Glass system predictor.** **a**. Reservoirs of varying sizes (N=20, 200, 1000) were setup to act as a unity signal follower (simplest test case) and tested on a MG system, which is a periodic but subtly chaotic time series. **b**. An example comparison between the training data set and the test sequence. **c**. Comparison between test and RC output signals. Reservoirs of all the sizes could reproduce the test signal. **d**. RC run in a generative mode, an example of a *temporal auto-encoder* where the network tries to generate the test signal looking only at its self-generated previous outputs. It can be seen that low N (one of the hyperparameters) reservoirs fail spectacularly, while higher N helps in generating a better fit.

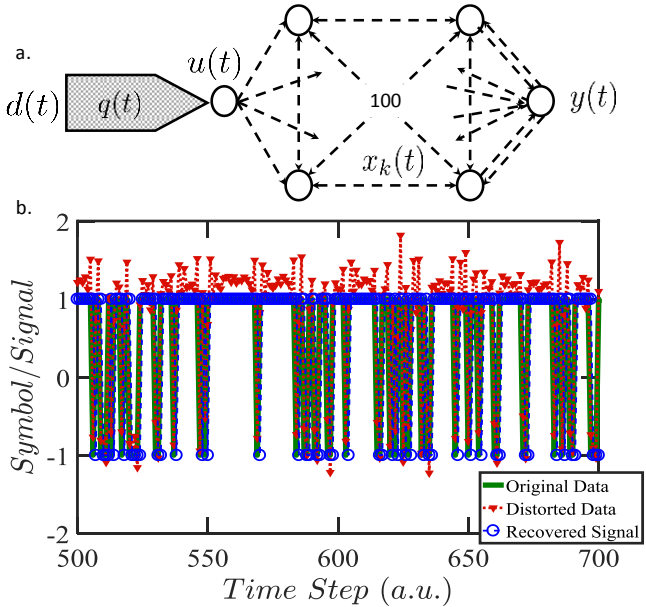

**Fig. 6. A Non-Linear Channel Equalizer. a**. Datastream d(t) is passed through a channel q(t) that causes non-linear distortions as well time shifts to the datastream. **b**. A N=10 RC can reproduce the original datastream d(t) with perfect fidelity from the distorted data u(t) (SER~0).

| Parameter | Units | Value |
|---|---|---|
| SPM U | kT | 1 (thermal limit) |
| SPM Volume | nm$^3$ | $\pi$x50x50x1 |
| GSHE θ | - | 0.33 |
| GSHE Volume | nm$^3$ | 50x50x2 |
| GSHE ρ | μΩ-cm | 170 |
| MTJ TMR | - | 100% |
| MTJ RA | Ω-μm$^2$ | 100 |
| R$^{up}$ | kΩ | 15 |
| Write current | μA | 15 |
| Vdd | V | 0.8 |

**Table.1. a.** Parameters chosen for various materials/components needed to build the p-bit.

| Component | Dynamic Power Dissipated (μW) |
|---|---|
| MTJ Reader | 10 (for reqd. Vdd) |
| GSHE Writer | 0.2 |
| Spintronic Total | 10.2 |
| R$^{up}$ | 10 (for reqd. Vdd) |
| Buffer (dual inverter design) | 150 |
| Silicon Total | 160 |
| Node Total | 170.2 |

**Table 1. b.** Dissipation in the various components of the p-bit. The dissipation is set primarily due to the requirements of the silicon components.